# Antiferromagnetic ordering and critical behavior induced giant magnetocaloric effect in distorted kagome lattice $Gd_3BWO_9$

Zhuoqun Wang[1#], Xueling Cui[1#], Tim Treu[2], Jiesen Guo[1], Xinyang Liu[1,3,4],
Marvin Klinger[2], Christian Heil[2], Nvsen Ma[1], Xianlei Sheng[1], Zheng Deng[3,5],
Xingye Lu[6], Xiancheng Wang[3,5], Wei Li[4, 7], Philipp Gegenwart[2], Changqing Jin[3,5]
and Kan Zhao[1*]

[1]*School of Physics, Beihang University, Beijing 100191, China*

[2]*Experimentalphysik VI, Center for Electronic Correlations and Magnetism, University of Augsburg, 86159 Augsburg, Germany.*

[3]*Beijing National Laboratory for Condensed Matter Physics, Institute of Physics, Chinese Academy of Sciences, Beijing 100190, China*

[4]*CAS Key Laboratory of Theoretical Physics, Institute of Theoretical Physics, Chinese Academy of Sciences, Beijing 100190, China*

[5]*School of Physical Sciences, University of Chinese Academy of Sciences, Beijing 100190, China*

[6]*Center for Advanced Quantum Studies, School of Physics and Astronomy, Beijing Normal University, Beijing 100875, China*

[7]*Peng Huanwu Collaborative Center for Research and Education, Beihang University, Beijing 100191, China*

\# These authors contributed equally to this work

*Corresponding author email: kan_zhao@buaa.edu.cn

# Abstract

We synthesize the high-quality $Gd_3BWO_9$ single crystal and investigate its low-temperature magnetic and thermodynamic properties. Below $T_N$ = 1.08 K, the anisotropic behavior of magnetic susceptibilities reveals that the $Gd^{3+}$ moments exhibit the dominant antiferromagnetic coupling along the *c*-axis, while displaying a ferromagnetic arrangement in kagome plane. With pronounced magnetic frustration, in adiabatic demagnetization refrigeration experiments starting from initial conditions of 9 T and 2 K, $Gd_3BWO_9$ polycrystal reaches a minimum temperature of 0.151 K, significantly lower than its $T_N$. Due to the high density of $Gd^{3+}$ ions ($S$=7/2), the maximum magnetic entropy change reaches over 50 J kg$^{-1}$ K$^{-1}$ under fields up to 7 T in $Gd_3BWO_9$, nearly 1.5 times as large as commercial sub-Kelvin magnetic coolant $Gd_3Ga_5O_{12}$(GGG). The *H-T* phase diagram of $Gd_3BWO_9$ under **H**//*c* exhibits field-induced critical behavior near the phase boundaries. This observation aligns with the theoretical scenario in which a quantum critical point acts as the endpoint of a line of classical second-order phase transitions. Such behavior suggests the importance of further investigations into the divergence of magnetic Grüneisen parameter in the vicinity of critical field at ultralow temperatures.

# Introduction

Geometrically frustrated magnets are typically characterized by competing interactions of localized spins that provide a unique platform for exploring exotic quantum magnetic states. Typical examples are antiferromagnets where transition metal ions or 4f rare-earth ions form the kagome lattice, giving rise to strong electron correlations, spin-orbit coupling, and/or magnetic interaction [1-3]. In recent years, various antiferromagnets based on kagome lattice and its modifications have been extensively studied, including both insulators and metals, such as the quantum spin liquid candidates $ZnCu_3(OH)_6Cl_2$ [4], quantum antiferromagnet $BaCu_3V_2O_8(OH)_2$ [5], tripod-kagome-lattice dipolar magnet $Dy_3Sb_3Zn_2O_{14}$ [6-7], distorted kagome lattice heavy fermion system CePdAl [8-11] and kagome spin ice HoAgGe [12-14], and three-dimensional hyperkagome-lattice quantum spin liquid candidate $PbCuTe_2O_6$ [15-16].

Quite recently, the rare-earth-based systems $R_3BWO_9$ (R=Pr, Nd, Sm, Gd, Tb, Dy, Ho) have attracted considerable attention as searching for new kagome antiferromagnets [17]. These compounds crystallize in a hexagonal coordinated structure with space group $P6_3$ (no. 173), where the $R^{3+}$ ions are connected by oxygen ions and form a distorted kagome lattice in the *ab*-plane [Fig. 1(a)] and stack in an AB-type arrangement along the *c*-axis [Fig. 1(b)]. Notably, the interlayer separations between $R^{3+}$ ions are slightly smaller than the intralayer distances [Fig. 1(b)], which enhances the three-dimensional nature of the magnetic interaction in $R_3BWO_9$ [17]. The non-Kramers system $Pr_3BWO_9$ has been identified as a frustrated quantum Ising magnet [18-19]. The Kramer system $Nd_3BWO_9$ exhibits short-range correlation below 1 K and enters the long-range magnetic order state below 0.3 K, accompanied by anisotropic field-induced metamagnetic quantum criticalities. The rich magnetic behavior of $Nd_3BWO_9$ can be understood through Ising model composed of twisted triangular spin-tubes [20-22]. For $Sm_3BWO_9$, a magnetic phase transition occurs at 0.75 K, and nuclear magnetic resonance spectra reveal an incommensurate magnetic order [23]. The polycrystalline studies of $Gd_3BWO_9$ demonstrate a giant magnetocaloric effect (MCE) while the magnetic structure remains unclear [24-25]. Therefore, investigating the anisotropic properties of $Gd_3BWO_9$ single crystals is crucial for understanding the microscopic mechanism of the MCE.

In this article, based on the magnetic susceptibilities along *a*-and *c*-axis, we demonstrate that the $Gd^{3+}$ moments are aligned antiferromagnetically along the *c*-axis and ferromagnetically in *ab*-plane below $T_N = 1.08$ K. Below $T_N$, the metamagnetic transitions of $Gd_3BWO_9$ single crystal are identified through magnetization and thermodynamic measurement under **H**//*c*. We continue to construct a phase diagram of magnetic entropy as a function of temperature and magnetic field along the *c*-axis, revealing a giant MCE at low temperature. Our work on $Gd_3BWO_9$ single crystal offers significant insights into its microscopic magnetic structure, magnetic critical behavior, and the associated giant MCE.

## Experiment

The polycrystalline Gd$_3$BWO$_9$ is synthesized using standard solid-state reaction method by mixing stoichiometric amounts of pre-dried Gd$_2$O$_3$, H$_3$BO$_3$ and WO$_3$. The powders are ground together and sintered in air at 1200 °C for 36 h, with several intermittent regrindings. High-quality single crystals of Gd$_3$BWO$_9$ were grown via flux method analogous to Ref. [26]. Yellow transparent single crystals with six hexagonal facets are obtained and have a typical mass ranging from a few milligrams to 100 mg [inset of Fig. 1(d)]. The X-ray diffraction (XRD) patterns and rocking-curve scans are collected using a Bruker D8 ADVANCE diffractometer with Cu-Kα radiation (λ = 1.5406 Å) at room temperature. Crystal orientation verification is performed using a Laue X-ray alignment system (Photonic Science Ltd.). The magnetic susceptibility and magnetization are measured as a function of the applied field (0 to 7 T) and temperature (0.4 to 300 K) using a Magnetic Property Measurement System SQUID magnetometer (MPMS, Quantum Design) equipped with a $^3$He-refrigerator insert. Specific heat measurements are performed using a Physical Property Measurement System (PPMS, Quantum Design) under applied magnetic fields of 3 T and 7 T with a $^3$He-option, and in zero field (0 T) with a $^3$He-$^4$He dilution refrigerator.

For the adiabatic demagnetization refrigeration (ADR) experiment in PPMS, we used a 4.38 g cylindrical pellet of 15 mm diameter and 4 mm thickness containing equal weights of Gd$_3$BWO$_9$ and silver powder. The pressed pellet was sintered at 800°C to further improve the thermal conductivity. The sample temperature was measured using a custom-built thermometer based on a commercial ruthenium oxide chip resistor. This thermometer was calibrated against a known reference thermometer and read out with a Lake Shore Model 372 AC resistance bridge, equipped with a Model 3726 scanner, operating at a constant current of 1 nA.

## Result and Discussion

### A. Crystal structure

Fig. 1(c) shows the room-temperature powder X-ray diffraction pattern and Rietveld refinement of Gd$_3$BWO$_9$. The observed profile aligns well with the simulation, yielding refinement reliability factors $R_p$=5.12% and $R_{wp}$=7.03%. The structural parameters determined from the Rietveld profile analysis are $a$ = 8.564(1) Å and $c$ = 5.402(1) Å, consistent with previous reports [17,25]. The distances between Gd and Gd in the kagome plane are 4.212 Å and 4.825 Å [Fig. 1(a)], while those in adjacent kagome planes are 3.880 Å and 4.311 Å [Fig. 1(b)]. The four comparable distances highlight the three-dimensional characteristics of the magnetic interaction. The complete crystal data are provided in the Supplemental Material [27].

Moreover, we have grown large-size single crystals by flux method, with typical crystals shown in the left inset of Fig. 1(d). The XRD patterns of the *bc*-plane [Fig. 1(d)] display dominant ($H$00) reflections. Rocking curve analysis of the Bragg peak (200) [right inset of Fig. 1(d)] demonstrates a narrow full-width-at-half-maximum (FWHM) of 0.11°, indicating high quality of the single crystal. As illustrated in Fig. 1(e-f), the six-fold symmetry of Laue pattern agrees well with the experimental simulation data, based on the crystal structure presented in Fig. l(a-b).

### B. Magnetic susceptibility and Magnetization

Fig. 2(a) presents the temperature dependence of magnetic susceptibilities $\chi(T)$ and inverse susceptibilities $1/\chi(T)$ for **H**//*a* and **H**//*c*, measured from 0.4 K to 300 K under 0.05 T. For low-temperature region (4-15 K) in Fig.2(b), the Curie-Weiss temperature is determined to be -0.6 K for **H**//*a* and -0.4 K for **H**//*c*, revealing the competition between predominant antiferromagnetic (AFM) interaction in adjacent layers and relatively weak intralayer ferromagnetic (FM) interaction within triangular framework of $Gd_3BWO_9$ [inset of Fig. 2(b)]. For high-temperature regime (200-300 K), the effective moment is calculated to be 7.82 $\mu_B$ for both field orientations, close to the expected value of 7.94 $\mu_B$.

The low-temperature $\chi(T)$ [Fig. 2(b)] shows an obvious peak centered at $T_N$ = 1.08 K, signaling long-range AFM ordering. Above $T_N$, isotropic susceptibility behavior reflects the absent orbital component of $Gd^{3+}$ moments. Below $T_N$, the $\chi(T)$ increases with the temperature decreasing under **H**//*a*, whereas it decreases under **H**//*c*. This anisotropic behavior demonstrates characteristic AFM ordering along the *c*-axis. As mentioned above, the nearest interlayer distance between $Gd^{3+}$ ions significantly enhances the AFM interaction along the *c*-axis, serving as the primary reason for magnetic ground state properties. Our experimental results provide direct evidence supporting the magnetic structure [inset of Fig. 2(b)] predicted by density functional theory (DFT) [24]. Similar phenomena have been observed in $Eu^{2+}$ containing $EuMnBi_2$ [28] and $EuMnSb_2$ [29]. In $EuMnBi_2$, the Eu moments order ferromagnetically in *ab* plane and align along *c* axis in the sequence of up-up-down-down below $T_N$~22 K, where the $\chi(T)$ parallel to the *c*-axis steeply decreases while it slightly increases vertical to the *c*-axis [28].

Fig. 2(c-d) depicts the isothermal magnetization curves $M(H)$ at various temperatures, and the magnetic moment is derived to be 6.85 $\mu_B$ for **H**//*a* and 6.78 $\mu_B$ for **H**//*c* at 0.4 K under 7 T, slightly smaller than the theoretical value 7 $\mu_B$. The $M(H)$ data can be well fitted to Brillouin function above $T$ = 5 K, which reveals the paramagnetic coupled nature between Heisenberg-like spins of $Gd^{3+}$ ions. Below 5 K, the deviation of the fitted curve at 3 K and 1.8 K indicates the appearance of short-range spin correlations.

To further analyze magnetic property of Gd$_3$BWO$_9$, Fig. 2(e-f) display the field derivative of magnetization d$M$/d$H$ curves at 0.4 K. For the d$M$/d$H$ curve under **H**//$a$ in Fig. 2(f), a pronounced peak is observed at 0.58 T, where the magnetization reaches 3.50 $\mu_B$, close to 1/2 of the saturated value ($M_s^a$). Additionally, a shoulder feature is present at 0.96 T, with the $M$ value of 5.10 $\mu_B$ roughly 3/4 $M_s^a$. In the case of **H**//$c$ [Fig. 2(e)], a sharp peak appears at 0.40 T, followed by a shoulder at 0.82 T. The respective $M$ values are 2.17 $\mu_B$ and 4.60 $\mu_B$, corresponding to approximately 1/3 and 2/3 of the saturated value ($M_s^c$), respectively. It would be valuable to investigate which variations in magnetic order states are associated with these anomalies in the d$M$/d$H$ curves. Remarkably, the magnetization curve of Nd$_3$BWO$_9$ exhibits a 1/3 fractional plateau for **H**//$c$ in the AFM order state, where the d$M$/d$H$ curve approaches zero in the vicinity of plateau region [20-22].

## C. Specific heat, Magnetic entropy and Magnetocaloric effect

To study thermodynamic properties of Gd$_3$BWO$_9$, the specific heat $C_p(T)$ measurements are performed down to 0.05 K in zero field, and 0.39 K in applied fields of 3 T and 7 T, as depicted in Fig. 3(a). The zero-field specific heat data displays a $\lambda$-shaped anomaly at $T_N$, consistent with the magnetic result above. The peak shape in $C_p(T)$ changes to a Schottky-type anomaly in an applied field of 3 T and 7 T.

Below $T_N$, the specific heat mainly comes from the magnetic contribution. As shown in inset of Fig. 3(a), the zero-field $C_{mag}(T)$ from 0.05 K to 0.2 K is fitted to a power law expression: $C_{mag}(T) \sim T^\gamma$, yielding $\gamma = 2.39(3)$, lower than the value of 3 typically observed in conventional gapless antiferromagnets [30]. In the triangular antiferromagnet KBaGd(BO$_3$)$_2$, the power law relationship with an exponent $\gamma = 2.0$ reflects the two-dimensional (2D) magnetic interaction intrinsic to the system [31-33]. In the pyrochlore Heisenberg antiferromagnet Gd$_2$Sn$_2$O$_7$, specific heat and inelastic neutron scattering (INS) measurements reveal the presence of magnon excitations with a gap ~ 1.5 K [34-36]. According to INS, the spin-spin correlations begin to develop around 20 K, and eventually form a gapped long-range ordered 'Palmer Chalker' state below $T_C = 1$ K. Below 0.3 K, the $C_{mag}(T)$ drops exponentially, and its value drops to 0.02 J mol$_{Gd}^{-1}$ K$^{-1}$ near 0.1 K [35]. In contrast, the $C_{mag}(T)$ of Gd$_3$BWO$_9$ displays a value around 0.2 J mol$_{Gd}^{-1}$ K$^{-1}$ at 0.1 K, an order of magnitude larger than that of Gd$_2$Sn$_2$O$_7$. Furthermore, the $C_{mag}(T)/T$ approaches a constant value of 4.3 J mol$^{-1}$ K$^{-2}$ below 0.07 K [Fig. 3(b)], indicating the unconventional gapless low-energy excitations in Gd$_3$BWO$_9$.

The lattice contribution arising from thermal vibration dominates the high temperature region. As shown in Fig. S1, an analysis is performed by fitting the zero-field specific heat data using a polynomial expression $C_{ph} = \beta_1 T^3 + \beta_2 T^5 + \beta_3 T^7 + \ldots$ [24], between 20 K and 50 K. The magnetic specific heat is obtained by subtracting the lattice contributions $C_{mag} = C_p - C_{ph}$. As temperature decreases, the zero-field $C_{mag}(T)/T$ curve of Gd$_3$BWO$_9$ begins to increase below 5 K and reaches its maximum value at $T_N$ [Fig. 3(b)]. The magnetic entropy $S_{mag}(T)$ is obtained by integrating $C_{mag}(T)/T$ data [Fig. 3(b)], and the value of $S_{mag}(T)$ is slightly larger than $3Rln8$ at 20 K in zero field [Fig. 3(c)].

To estimate the minimum temperature in ADR, the adiabatic demagnetization process is roughly depicted by arrows in Fig. 3(c) [31-32]. The $S_{mag}(T)$ is reduced by 44.02 J mol$^{-1}$ K$^{-1}$ when applying a field of $\mu_0 H = 7$ T from an initial temperature $T_0 = 2$ K (isothermal). Driving the field back to 0 T adiabatically is supposed to cool the material to 0.21 K. When initiating from $T_0 = 5$ K under 7 T, the material could cool down to 0.60 K with the $S_{mag}(T)$ reduction of 37.33 J mol$^{-1}$ K$^{-1}$. Similarly, the commercial magnetic coolant Gd$_3$Ga$_5$O$_{12}$ (GGG) achieves a minimum temperature of 0.32 K with an initial condition of $T_0 = 2$ K and $\mu_0 H = 6$ T [33,37], and another spin chain containing antiferromagnet NaGdP$_2$O$_7$ achieves a minimum temperature of 0.22 K with $T_0 = 2$ K and $\mu_0 H = 5$ T [38].

We continue to directly test the refrigeration performance of Gd$_3$BWO$_9$ polycrystal using the modified ADR cooling setup reported in Ref. [32,38-39]. Starting from an initial condition 2 K and 9 T, the sample temperature instantaneously decreases as soon as the field begins to decrease. Upon ramping the field down to zero at a sweep rate of 5 mT/s, a local minimum $T_{min2} = 0.156$ K is achieved at $\mu_0 H = 0.90$ T, and another minimum $T_{min1} = 0.151$ K is achieved at $\mu_0 H = 0.22$ T in Run 1 [Fig. 3(d)], consistent with the estimated value 0.21 K under 2 K and 7 T condition. We will discuss the temperature minimum and phase boundary later. The temperature starts to increase from $T_0 = 0.164$ K as magnetic field drops to zero [inset of Fig. 3(d)]. As depicted in Fig.3(e), the warm-up time from $T_0$ up to 0.48 K exceeded 24 hours in Run 1, indicating exceptional holding time.

To calculate the specific heat of Gd$_3$BWO$_9$ sample from ADR, we conduct a second run with a resistance heater on the pellet. As shown in Fig. 3(e), there seems a slope change in the warming curve around $T_N = 1.08$ K, due to the AFM ordering of Gd$_3$BWO$_9$. As the external resistive heat input was accurately measured and found to be nearly constant as $\dot{Q} = 1.4$ µW, much larger than the natural heat leakage into the system, which is below 100 nW, the slope of the warming curve is closely associated with the sample's heat capacity. According to $\dot{Q} = C_{ADR}\dot{T}$, $C_{ADR}$ obtained in Run 2 is good agreement with the 'directly' measured heat capacity [Fig. 3(f)].

In order to gain deeper insights into the MCE of Gd$_3$BWO$_9$, the magnetic entropy change -$\Delta S_{\text{mag}}$ is calculated based on the $M(H)$ curves in Fig. S2, through Maxwell's thermodynamic relation -$\Delta S_{\text{mag}}$ = -$\int_0^H (\frac{\partial M}{\partial T})_{P,H} dH$ [24]. For each curve, we have used two sets of $M(H)$ data at temperatures $T$-$\Delta T$ and $T$+$\Delta T$. Through $\partial M/\partial T \approx \Delta M/\Delta T = [M(T+\Delta T, H)-M(T-\Delta T, H)]/2\Delta T$ [Fig. 4(a-b)], the field dependence of the entropy -$\Delta S_{\text{mag}}(H)$ could be obtained by integrating $\Delta M/\Delta T$ over the field range [38,40]. For instance, the -$\Delta S_{\text{mag}}(H)$ of 0.7 K is calculated from the $M(H)$ of 0.4 K and 1 K.

The field variation of -$\Delta S_{\text{mag}}$ is shown in Fig. 4(c) for **H**//$a$ and Fig. 4(d) for **H**//$c$, respectively. The contour plots of -$\Delta S_{\text{mag}}$ for **H**//$c$ [Fig. 4(e)] are similar to the polycrystalline data [24-25]. According to the -$\Delta S_{\text{mag}}$ for **H**//$c$, the temperature dependence of -$\Delta S_{\text{mag}}$ is depicted under different fields in Fig. 4(f), and the -$\Delta S_{\text{mag}}(T)$ derived from $C_{\text{mag}}(T, H)$ data closely matches the -$\Delta S_{\text{mag}}(T)$ curves calculated from $M(H, T)$ data. Given the high spin state and high density of Gd$^{3+}$ ions, the theoretically calculated -$\Delta S_{\text{mag}}$ reaches 51.87 J mol$^{-1}$ K$^{-1}$ (63.99 J kg$^{-1}$ K$^{-1}$) for Gd$_3$BWO$_9$, and the volumetric entropy density reaches 483 mJ K$^{-1}$ cm$^{-3}$, surpassing most Gd-oxide magnets, including GGG (363 mJ K$^{-1}$ cm$^{-3}$), GdPO$_4$ (401 mJ K$^{-1}$ cm$^{-3}$), and KBaGd(BO$_3$)$_2$ (192 mJ K$^{-1}$ cm$^{-3}$), while remaining comparable to Gd$_{9.33}$[SiO$_4$]$_6$O$_2$ (509 mJ K$^{-1}$ cm$^{-3}$) [41]. During our experiment, the -$\Delta S_{\text{mag}}$ achieves the maximum value 43.32 J mol$^{-1}$ K$^{-1}$ (53.45 J kg$^{-1}$ K$^{-1}$) and 42.82 J mol$^{-1}$ K$^{-1}$ (52.83 J kg$^{-1}$ K$^{-1}$) at 2.1 K under field up to 7 T along $a$ and $c$ axes, accounting for 83.5% and 82.6% of the theoretical value, respectively. These values exceed the maximum value of GGG (38.4 J kg$^{-1}$ K$^{-1}$) at 2 K under a 7 T field [42-43]. The large volumetric entropy density, magnetic entropy change, and low temperature in ADR suggest that Gd$_3$BWO$_9$ holds significant potential for refrigeration applications.

### D. Phase diagram and Magnetic critical behavior under H//c

More interestingly, at $T$ = 0.7 K< $T_N$, the value of -$\Delta S_{\text{mag}}$ is negative when the magnetic field is less than 1 T for **H**//$c$, while it is always positive for **H**//$a$ [Fig. 4(c-d)]. This anisotropic response can be attributed to the dominant AFM coupling along the $c$-axis between Gd$^{3+}$ ions. The emergence of two anomalies in the d$M$/d$H$ curve under $\mu_0 H$ = 0.40 T and 0.82 T at $T$ = 0.4 K may be associated with critical behavior in thermodynamic properties. Below, we map the phase diagram for **H**//$c$ with more detailed magnetic and thermodynamic data.

As shown in Fig. 5(a), the low temperature $\chi(T)$ curves of Gd$_3$BWO$_9$ are measured under several small fields along $c$-axis, whose peak shifts to lower temperature as field increases. In addition, we conducted the $M(H)$ measurements with more data points below 3 T at 0.4 K, 0.6 K, 0.8 K, 1.0 K and 1.2 K, respectively [Fig. S3] [27]. Correspondingly, the Fig. 5(c-d) and Fig. S4 show low field variation of $\Delta S_{\mathrm{mag}}$ curves at various temperatures, calculated based on the $M(H)$ curves in Fig. S3 [27]. Since $\Delta S_{\mathrm{mag}}(H)=S_{\mathrm{mag}}(H)-S_{\mathrm{mag}}(0\,T)$, the $S_{\mathrm{mag}}(H)$ curve at a constant temperature can be obtained through vertically shifting the $\Delta S_{\mathrm{mag}}(H)$ curve by the zero-field entropy value, as shown in Fig.5(b). It is worth noting that a shoulder around $\mu_0 H_{c1} = 0.42$ T and a peak at $\mu_0 H_{c2} = 0.84$ T are observed for the $S_{\mathrm{mag}}(H)$ curve at 0.5 K, which gradually evolve into one single peak as temperature increases. Taken the features mentioned above, Fig. 5(e) displays an $H$-$T$ phase diagram of $Gd_3BWO_9$ under $\mathbf{H}//c$, with the color coding representing $S_{\mathrm{mag}}(H, T)$ in Fig. 5(b).

The isothermal field scans on $C_{\mathrm{mag}}(H)$ are measured under $\mathbf{H}//c$ at 0.9 K and 0.4 K, respectively. As shown in the top panel of Fig. 5(c-d), a peak is identified on $C_{\mathrm{mag}}(H)$ curve at $\mu_0 H = 0.5$ T at 0.9 K, while upon cooling to 0.4 K, a shoulder is observed around $\mu_0 H = 0.85$ T in addition to the peak at $\mu_0 H = 0.55$T. The magnetic Grüneisen parameter is calculated through $\Gamma_{\mathrm{mag}} \equiv \frac{1}{T}(\frac{\mathrm{d}T}{\mathrm{d}\mu_0 H})_S = -\frac{1}{\mu_0 C}\frac{\mathrm{d}S}{\mathrm{d}H} \approx -\frac{1}{\mu_0 C_{\mathrm{mag}}}\frac{\mathrm{d}S_{\mathrm{mag}}}{\mathrm{d}H}$, which the non-magnetic specific heat is much smaller than the magnetic component below $T_{\mathrm{N}}$. Combined with Maxwell's thermodynamic relation $(\frac{\partial S}{\partial H})_{P,T} = (\frac{\partial M}{\partial T})_{P,H}$, the $\Gamma_{\mathrm{mag}}(H)$ curves are shown in Fig. 5(c-d).

At 0.9 K, the sign of $\Gamma_{\mathrm{mag}}(H)$ curve changes from negative to positive at $\mu_0 H_c = 0.5$ T, indicating a local maximum in entropy $S_{\mathrm{mag}}(H)$ at this field, as well as a respective minimum in the adiabatic temperature trace $T(H)_S$, arising at the phase boundary. At $T = 0.4$ K, there is a local maximum in $\Gamma_{\mathrm{mag}}(H)$ curve at $\mu_0 H_{c1} = 0.42$ T, with its value approaching zero. As the magnetic field increases beyond $\mu_0 H_{c2} = 0.85$ T, the $\Gamma_{\mathrm{mag}}(H)$ curve changes its sign from negative to positive, reaching to a constant value up to 1.5 T. The lower critical field $\mu_0 H_{c1} = 0.42$ T may indicate a metamagnetic transition within the AFM regime, and the system subsequently reaches a fully polarized state above upper critical field $\mu_0 H_{c2} = 0.85$ T. This behavior is similar to that in $Nd_3BWO_9$, which a proximate quantum bicritical point (BCP) is observed along the $c$-axis. In the magnetocaloric effect results of $Nd_3BWO_9$, a temperature diplike feature appears near the critical fields $B_{\mathrm{BCP}} \sim 0.61$ T, where magnetic entropy change $\Delta S_{\mathrm{mag}}(H)$ accumulates its maximum value [21].

As detailed in Fig. 4 of Ref. [44], when the quantum critical point (QCP) serves as the endpoint of a line of classical second order transitions, the isentropes (d$S$=0), i.e. the contour lines in $S_{\mathrm{mag}}(H, T)$ would exhibit minima values close to the phase boundary. Intriguingly, the contour lines in $S_{\mathrm{mag}}(H, T)$ in Fig. 5(e) explicitly illustrate the evolution of temperature during an adiabatic sweep of the field along $c$ axis at $T > 0.4$ K. This observation aligns precisely with the light blue contour line depicted in Fig. 5(e). It also includes the adiabatic temperature traces of field sweep $T(H)_S$ of $Gd_3BWO_9$ polycrystal. These sweeps were measured by ramping the magnetic field from 9 T to 1.5 T, as in the ADR runs. Subsequently, the field was swept repeatedly at a substantially reduced rate of 1.5 mT/s from 1.5 T to 0 T and back to 1.5 T. Though the critical field might be slightly different from the polycrystal compared with single crystal under **H**//$c$ axis, the temperature traces $T(H)_S$ fit quite nicely into the phase diagram based on $S_{mag}(H, T)$. By drawing boundary lines through local minima of the temperature traces, the phase boundaries could be extended to lower temperatures.

For a quantum-critical point (QCP), there shall be the sign change at the critical field $B_c$ and also a divergence $\Gamma_{mag} \sim |B-B_c|^{-1}$ at very low temperature. Thus, the quantum critical phenomena in $Gd_3BWO_9$ at $T < 0.1$ K, including the enhanced MCE approaching the putative QCPs warrant further investigation in the future [40, 44-47].

## Summary

In summary, we have performed magnetic and thermodynamic characterization of distorted kagome lattice $Gd_3BWO_9$ single crystal. Below $T_N = 1.08$ K, the $Gd^{3+}$ moments exhibit AFM alignment between layers and FM alignment within layers, resulting in alternating AFM-FM stripes along the $c$-axis, as confirmed by magnetic susceptibility measurements. Compared to the GGG case, the maximum $-\Delta S_{mag}$ of $Gd_3BWO_9$ shows larger values of 53.45 J kg$^{-1}$ K$^{-1}$ for **H**//$a$ and 52.83 J kg$^{-1}$ K$^{-1}$ for **H**//$c$ under 7 T, presenting the giant MCE related to its AFM order. Given the long-range order at $T_N = 1.08$ K, the strong magnetic frustration of $Gd_3BWO_9$ suppresses the release of magnetic entropy down to low temperature, resulting in a $T_{min} = 0.151$ K far below $T_N$ in the ADR measurement. Furthermore, the constructed $H$-$T$ phase diagram of $Gd_3BWO_9$ under **H**//c reveals evidence of magnetic critical behavior, as indicated by the sign change of $\Gamma_{mag}(H)$ near the metamagnetic transitions.

## Acknowledgements

The authors would like to thank Peijie Sun, Zhaoming Tian, Junsen Xiang, Huifen Ren and Shaokui Su for helpful discussions and experimental support. The work was supported by the National Key R&D Program of China (Grant No. 2023YFA1406003),

National Natural Science Foundation of China (Grants No. 12274015), the Beijing Natural Science Foundation (Grant No. JQ24012), and the Fundamental Research Funds for the Central Universities. The work in Augsburg was supported by the Deutsche Forschungsgemeinschaft (DFG, German Research Foundation), Grants No. 514162746 (GE 1640/11-1) and No. TRR 360-492547816. The authors acknowledge the facilities, and the scientific and technical assistance of the Analysis & Testing Center, Beihang University. A portion of this work was carried out at the Synergetic Extreme Condition User Facility (SECUF).

## Data Availability

The data that supports the findings of this article are available upon reasonable request from the authors.

# Figures and Captions

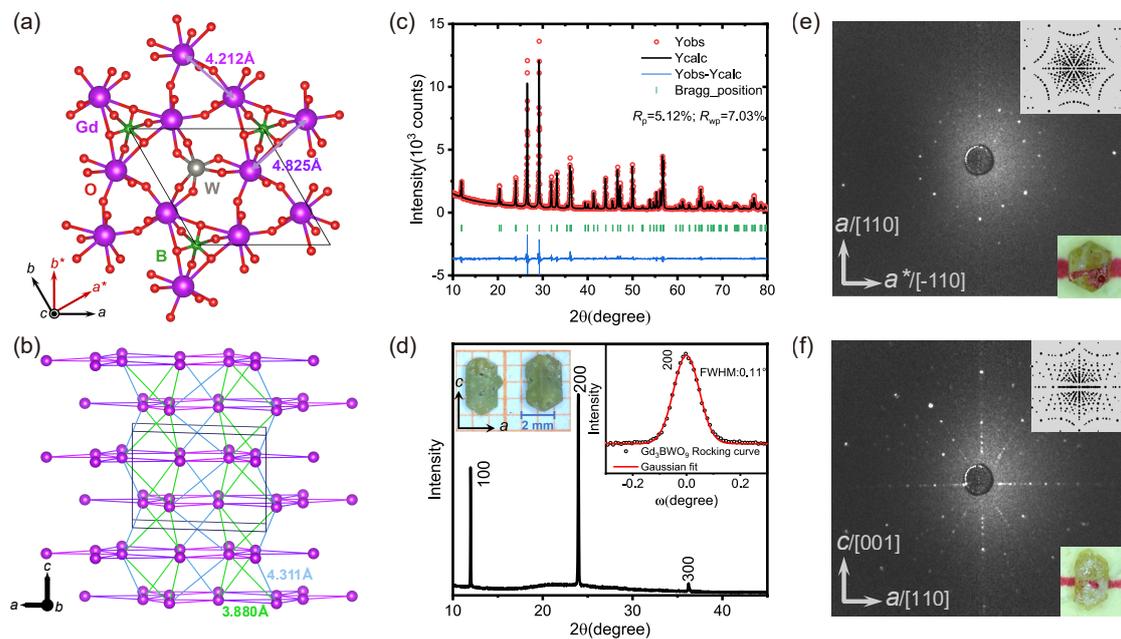

Fig. 1: **Crystal structure of $Gd_3BWO_9$.** (a) The schematic structure of $Gd_3BWO_9$ reflecting the distorted kagome lattice in the crystallographic ab plane. (b) The well-separated distorted kagome planes are stacked along the $c$ axis of $Gd_3BWO_9$. (c) The experimental XRD patterns of $Gd_3BWO_9$ powder. (d) X-ray diffraction pattern of $Gd_3BWO_9$ single crystal. Left inset: An optical image of typical single crystal. Right inset: the rocking curve (black line) of the (200) Bragg peak, together with the Gaussian fitting (red line) of $Gd_3BWO_9$. (e) and (f) The experimental and simulated Laue diffraction patterns of the $bc$ and $ac$ plane of $Gd_3BWO_9$, respectively.

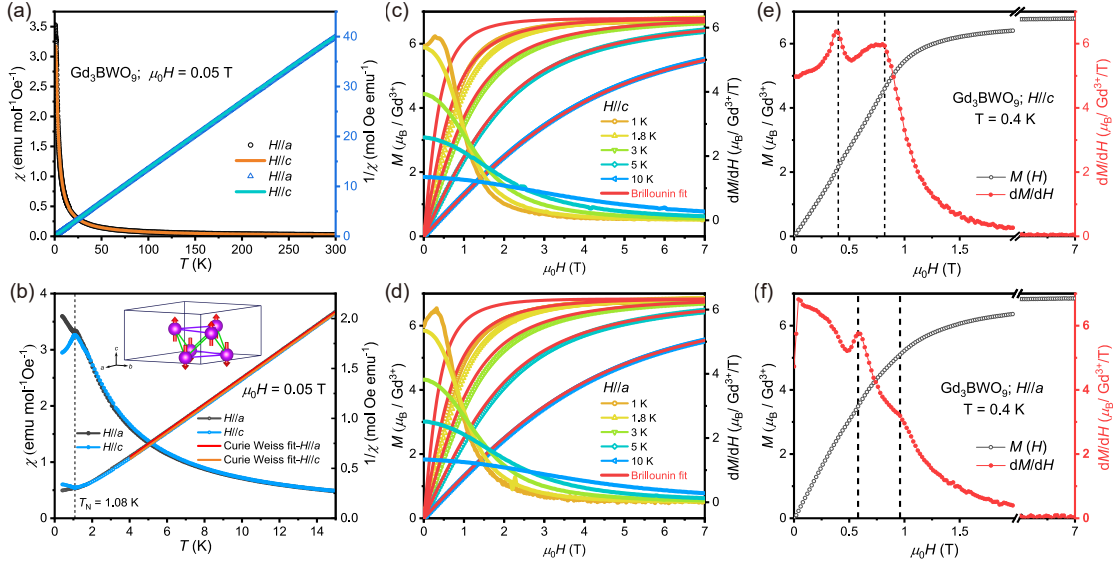

Fig. 2: **Magnetic properties of Gd$_3$BWO$_9$ under H//a and H//c.** (a) The susceptibility $\chi(T)$ and the inverse magnetic susceptibility [$1/\chi(T)$] of Gd$_3$BWO$_9$ for **H**//a and **H**//c under 500 Oe. (b)The low-temperature part of (a). The insert of (b): A schematic of the magnetic structure in Gd$_3$BWO$_9$, the green and purple lines constitute triangles connecting Gd ions in adjacent kagome plane (3.880 Å) and within kagome plane (4.825 Å). (c)-(d) Isothermal magnetization $M(H)$ and the derivative [d$M(H)$/d$H$] curves for selected temperatures of 1, 1.8, 3, 5, and 10 K along $c$ and $a$-axis. The solid red lines show the Brillouin function fits. (e)-(f) $M(H)$ and d$M(H)$/d$H$ curves along $c$ and $a$-axis at 0.4 K. The dotted lines represent the features.

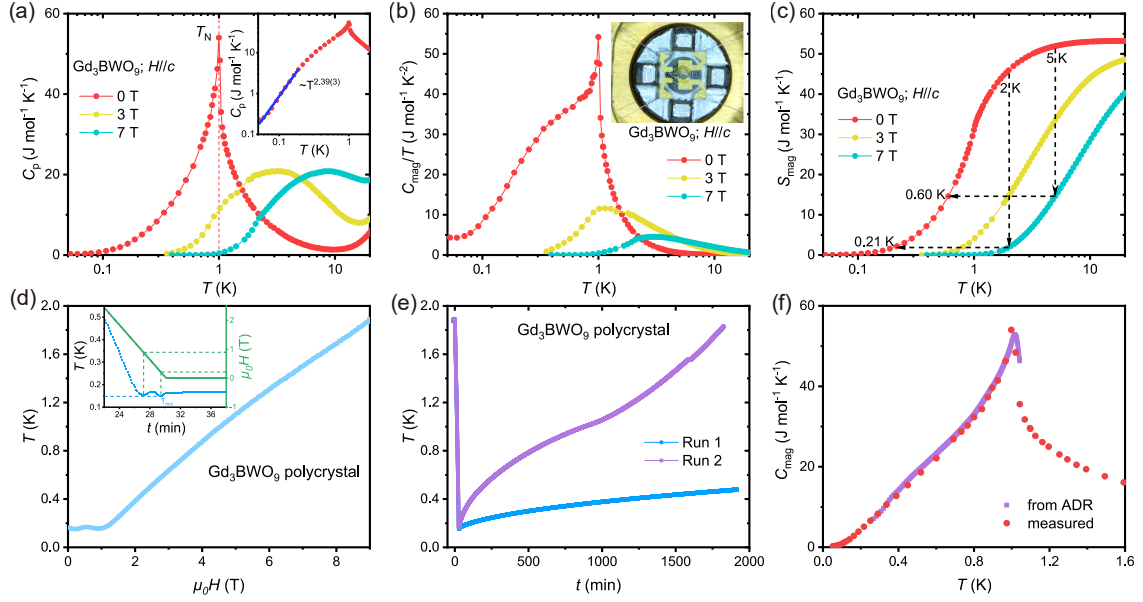

Fig. 3: **Specific heat result of Gd$_3$BWO$_9$ along H//c and ADR of Gd$_3$BWO$_9$ polycrystal.** (a) Total specific heat $C_p(T)$ measured down to 0.05 K in zero field, and 0.39 K in applied fields 3 T and 7 T along **H**//$c$. The inset shows a power law fit from 0.05 to 0.2 K in zero field. (b) Magnetic specific heat as $C_{mag}(T)/T$ in a logarithmic presentation at 0, 3, and 7 T. The inset shows a photo of the sample on the platform of the specific-heat puck. (c) Temperature variation of the magnetic entropy $S_{mag}$ computed from the heat capacity data of Gd$_3$BWO$_9$ for **H**//$c$. The arrows designate the adiabatic demagnetization process starting from initial temperatures of 2 K and 5 K under 7 T. (d) The cooling curve of Gd$_3$BWO$_9$ polycrystal in the ADR process starting from 9 T and 2 K (slightly shifted due to the neglect of the magnetoresistance of the thermometer). The inset shows the process around $T_{min}$. (e) The temperature evolution was measured over a period exceeding 30 hours, with Run 1 conducted in the absence of a resistance heater. (f) The heat capacity calculated from the warming curve of the ADR experiment and directly measured in the PPMS.

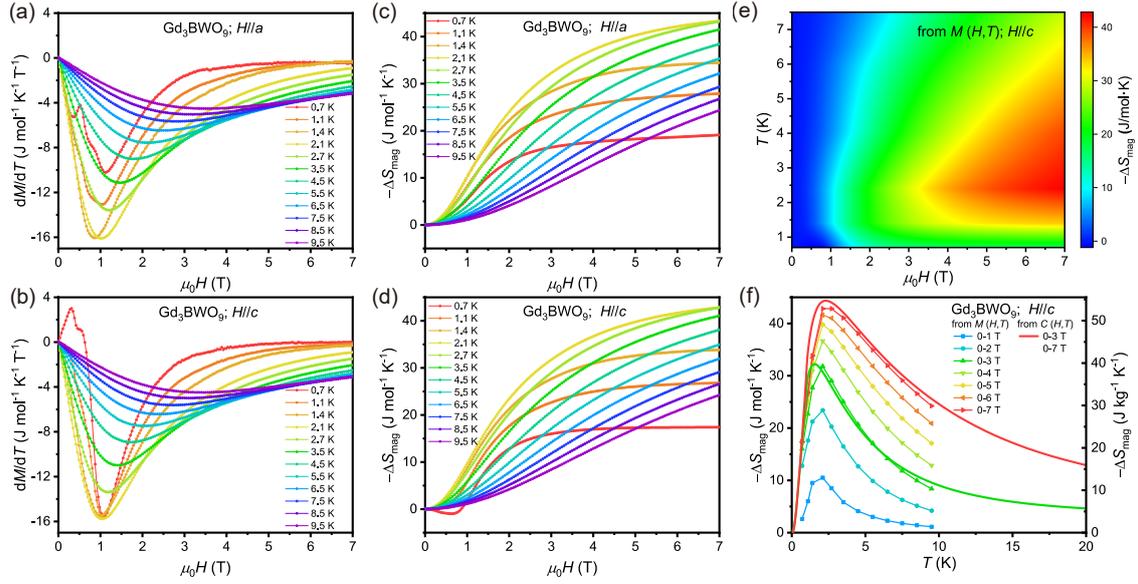

Fig. 4: **Magnetic entropy and Magnetocaloric effect analysis of Gd$_3$BWO$_9$.** (a)-(b) Derivative of the magnetization with respect to temperature as a function of the applied magnetic field for **H**//*a* and **H**//*c*. (c)-(d) Field variation of the magnetic entropy change -Δ$S_{mag}$ calculated from isothermal magnetization curves for different temperature of Gd$_3$BWO$_9$ for **H**//*a* and **H**//*c*, respectively. (e) Contour plots for -Δ$S_{mag}$ as a variation of applied field and temperature. (f) Temperature variation of -Δ$S_{mag}$ obtained from isothermal magnetization curves and heat capacity data, with vertical axes representing units in volumetric J mol$^{-1}$ K$^{-1}$ (left) and J kg$^{-1}$ K$^{-1}$ (right).

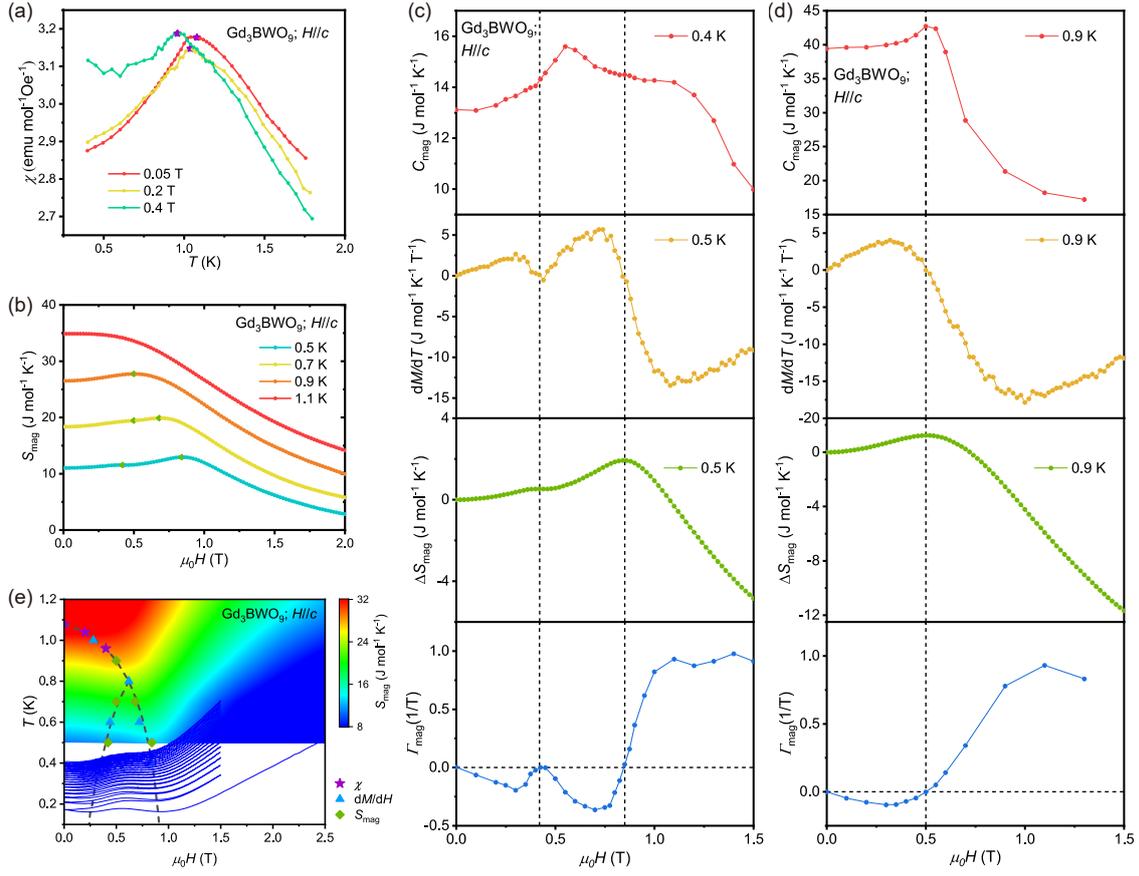

Fig. 5: **The phase diagram of Gd$_3$BWO$_9$ single crystal under H//c.** (a) The low-temperature susceptibility $\chi(T)$ curves under 0.05, 0.2, and 0.4 T for **H**//$c$, the purple stars indicate the positions of peaks. (b) The low field variation of magnetic entropy $S_{\text{mag}}(H)$ for **H**//$c$ at 0.5, 0.7, 0.9, and 1.1 K, respectively; the green rhombuses denote the positions of local peaks. The magnetic specific heat $C_{\text{mag}}$, the derivative of magnetization with respect to temperature d$M$/d$T$, magnetic entropy change $\Delta S_{\text{mag}}$, magnetic Grüneisen parameter $\Gamma_{\text{mag}}$ at 0.4 K (the $C_{\text{mag}}$ and $\Gamma_{\text{mag}}$ of 0.4 K would be similar with that of 0.5 K) (c) and 0.9 K (d) under **H**//$c$. (e) The $H$–$T$ phase diagram of Gd$_3$BWO$_9$ as derived from $\chi(T)$ (purple stars), d$M$/d$H$ (blue triangles), $S_{\text{mag}}(H)$ (green rhombuses), and $T(H)_S$ (blue lines), with the dashed lines guiding the eyes and the color coding representing $S_{\text{mag}}(H)$.